\documentclass[letterpaper]{article} 
\usepackage{aaai2026}
\usepackage{longtable}
\usepackage{tcolorbox}
\usepackage{fancyvrb}

\usepackage{times}  
\usepackage{helvet}  
\usepackage{courier}  
\usepackage[hyphens]{url}  
\usepackage{graphicx} 
\usepackage{natbib}  
\usepackage{caption} 
\usepackage{algorithm}
\usepackage{algorithmic}
\usepackage{enumitem}
\usepackage[percent]{overpic}
\usepackage{newfloat}
\usepackage{listings}
\DeclareCaptionStyle{ruled}{labelfont=normalfont,labelsep=colon,strut=off} 
\floatstyle{ruled}
\newfloat{listing}{tb}{lst}{}
\floatname{listing}{Listing}
\usepackage{amsmath}
\usepackage{amssymb}
\usepackage{multirow}
\usepackage{booktabs}
\usepackage{bbding}
\newcommand{\mname}{Chem3DLLM}

\nocopyright 
\title{\mname: 3D Multimodal Large Language Models for Chemistry}

\author{
\textbf{Lei Jiang\textsuperscript{1}}\quad
\textbf{Shuzhou Sun\textsuperscript{2,3}}\quad
\textbf{Biqing Qi\textsuperscript{2}} \quad
\textbf{Yuchen Fu\textsuperscript{2,4} }\quad
\textbf{Xiaohua Xu\textsuperscript{1,*} } \\
\textbf{Yuqiang Li\textsuperscript{2,*}} \quad
\textbf{Dongzhan Zhou\textsuperscript{2,*}}\quad
\textbf{Tianfan Fu\textsuperscript{4,2,*}}\\
\tt\small\textsuperscript{1}University of Science and Technology of China \\
\tt\small\textsuperscript{2}Shanghai Artificial Intelligence Laboratory \\
\tt\small\textsuperscript{3}CMVS, University of Oulu\quad
\textsuperscript{4}Nanjing University\\
\tt\small{jianglei0510@mail.ustc.edu.cn} 
\tt\small{shuzhou.sun@oulu.fi}
\tt\small{yuchenfu@smail.nju.edu.cn} \quad
\tt\small{xiaohuaxu@ustc.edu.cn\textsuperscript{*}}
\tt\small{\{qibiqing, liyuqiang\textsuperscript{*}, zhoudongzhan\textsuperscript{*}\}}@pjlab.org.cn\\
\tt\small{futianfan@gmail.com\textsuperscript{*}, *: correspondence}
}

\usepackage{bibentry}

\begin{document}

\maketitle

\begin{abstract}
In the real world, a molecule is a 3D geometric structure. Compared to 1D SMILES sequences and 2D molecular graphs, 3D molecules represent the most informative molecular modality. 
Despite the rapid progress of autoregressive-based language models, they cannot handle the generation of 3D molecular conformation due to several challenges: {1) 3D molecular structures are incompatible with LLMs' discrete token space, 2) integrating heterogeneous inputs like proteins, ligands, and text remains difficult within a unified model, and 3) LLMs lack essential scientific priors, hindering the enforcement of physical and chemical constraints during generation.} 
To tackle these issues, we present \mname, a unified protein-conditioned multimodal large language model. 
Our approach designs a novel reversible text encoding for 3D molecular structures using run-length compression, achieving 3× size reduction while preserving complete structural information. This enables seamless integration of molecular geometry with protein pocket features in a single LLM architecture. We employ reinforcement learning with stability-based rewards to optimize chemical validity and incorporate a lightweight protein embedding projector for end-to-end training. Experimental results on structure-based drug design demonstrate state-of-the-art performance with a Vina score of -7.21, validating our unified multimodal approach for practical drug discovery applications.
\end{abstract}
\section{Introduction}
\label{sec_Introduction}





In the real world, a molecule is a 3D geometric structure, characterized by its spatial arrangement of atoms and bonds. Compared to 1D SMILES sequences and 2D molecular graphs that provide valuable but limited representations, 3D molecules represent the most informative molecular modality~\citep{li2024towards,han2024generalist}.
For example, in structure-based drug design (SBDD) tasks, as the goal is to design a ligand that could bind tightly to the target protein, molecular spatial conformations directly determine compatibility and binding affinity with protein binding pockets~\citep{platzer2025nmr}. 
Therefore, three-dimensional molecular structure modeling represents a fundamental challenge in chemistry, drug discovery and materials science~\cite{huang2022artificial,zhang2023artificial}. 
\begin{figure}[ht]
\centering
\includegraphics[width=\columnwidth]{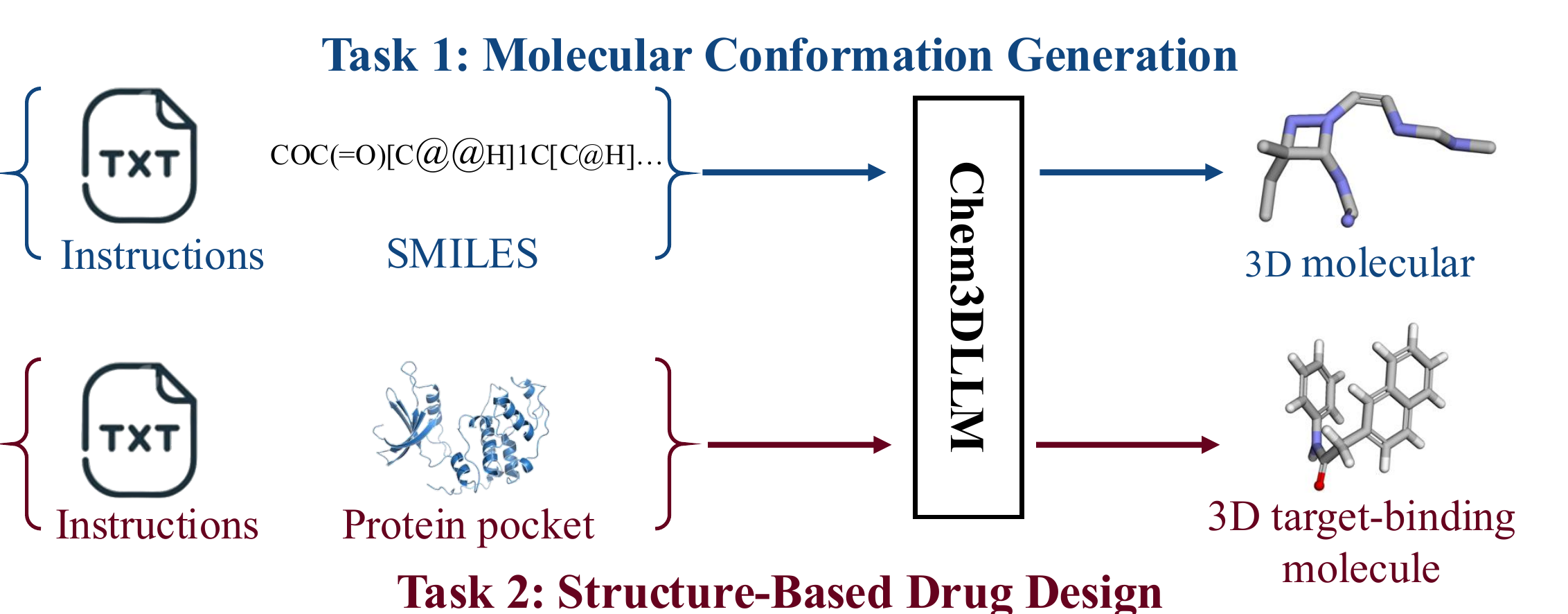} 
\caption{Chem3DLLM generates 3D molecules from different input modalities, such as SMILES strings and protein binding pockets.}
\label{fig:motivation}
\end{figure}

Large language models (LLMs) represent cutting-edge generative AI methods not only in the general-domain text generation but also in computational chemistry. In generative AI, LLMs have demonstrated remarkable capabilities in creating coherent text, generating creative content, and even writing code~\cite{achiam2023gpt4, team2024gemini, anthropic2024claude3, liu2023llava,team2024qwen2}. When applied to chemistry, LLMs offer transformative potential by enabling more accurate predictions of molecular properties, automating complex computational tasks, and enhancing the interpretation of experimental data~\cite{zhang2023moleculegpt,zhao2024chemdfm, zhang2024chemllm,li2025chemvlm,tan2025chemmllm}. Their ability to process and understand vast amounts of chemical information makes them invaluable tools for advancing research and development in materials science, drug discovery, and beyond~\citep{han2024generalist}. 

However, vanilla LLMs fail to generate 3D molecular structures directly due to the following challenges. 

\noindent\textbf{Challenge 1: The data format of 3D molecular structure is not compatible with LLM. } 3D molecular structures are typically represented as coordinate arrays (e.g., Cartesian coordinates of atoms) and interatomic distance matrices, which are numerical and continuous—formats fundamentally different from the discrete token sequences that LLMs are designed to process. Without appropriate encoding or discretization, these geometric data cannot be directly fed into or generated by standard language models, limiting their ability to reason about spatial configurations. 

\noindent\textbf{Challenge 2: Aligning multiple modalities (e.g., text, protein, ligand) in a unified model. } In real-world chemical applications, tasks often require joint understanding of heterogeneous inputs such as textual descriptions, protein structures, and small molecule ligands. Effectively aligning these diverse modalities in a shared embedding space while preserving their structural and semantic integrity remains a significant challenge for multimodal reasoning and generation. 

\noindent\textbf{Challenge 3: Incorporation of scientific prior knowledge. } Accurate 3D molecular generation requires adherence to physical and chemical constraints—such as bond lengths, angles, and stereochemistry—that are not naturally captured by LLMs trained on text corpora. Integrating this domain-specific scientific knowledge into the model’s architecture or training process is essential to ensure chemically valid and physically plausible outputs, but doing so in a scalable and generalizable way is non-trivial.

To address these three challenges, we propose \mname, a unified protein-conditioned three-dimensional molecular generation framework. \mname\ enables end-to-end modeling and optimization of 3D molecular conformations under the guidance of dual-conditional inputs: protein structures and molecular SMILES
as shown in Figure~\ref{fig:motivation}. Concretely, we propose the following three solutions. 

\noindent\textbf{Solution 1} (handling Challenge 1): we design Reversible Compression of Molecular Tokenization (RCMT), a novel reversible SDF-to-Text\footnote{An SDF (Structure Data File) is a chemical file format that stores molecular structures in 3D, including atomic coordinates, bond information, and associated molecular properties, often used for compound databases and cheminformatics applications. } compression mechanism that preserves three-dimensional geometric and chemical bond information while significantly reducing representation length.

\noindent\textbf{Solution 2} (handling Challenge 2): we introduce a lightweight protein structure projection module that maps spatial embedding features of protein pockets to the token semantic space of language models, forming a unified representational foundation with small-molecule encoding. 

\noindent\textbf{Solution 3} (handling Challenge 3): we propose Reinforcement Learning with Scientific Feedback (RLSF), a training paradigm that incorporates domain-specific physical and chemical priors—such as energy minimization, valency rules, and steric feasibility—as differentiable reward signals to guide the LLM’s generation process. By formulating 3D molecular conformation optimization as a sequential decision-making problem, RLSF enables the model to iteratively refine its outputs through feedback from a scientific critic module, which evaluates structural validity, binding complementarity, and energetic plausibility, thereby closing the loop between generative modeling and domain-knowledge enforcement. 

Our approach establishes a structure-aware generation pathway of ``representation → alignment → optimization'', providing comprehensive support for high-quality molecular design in three-dimensional space. 
Notably, unlike existing methods that treat molecular generation and protein-ligand interactions as independent tasks, our approach can jointly learn molecular conformation generation (MCG) and structure-based drug design (SBDD).

\noindent\textbf{Main contributions.}
Our main contributions are summarized as follows:
\begin{itemize}
\item We propose Reversible Compression of Molecular Tokenization (RCMT), a novel method that enables lossless compression of 3D molecular structures from SDF format into compact text sequences, preserving both geometric coordinates and chemical bond information while making the data compatible with language model processing. 
\item We introduce a lightweight protein structure projection module that aligns 3D protein pocket features with the semantic space of language models, enabling unified multimodal representation learning with small-molecule encodings. 
\item We propose Reinforcement Learning with Scientific Feedback (RLSF), a training framework that uses rewards based on physical and chemical principles to guide an LLM in generating valid and plausible 3D molecular conformations. By leveraging a scientific reward to provide feedback on structural and energetic feasibility, RLSF enables iterative refinement of molecular structures, effectively integrating domain knowledge into the generative process.
\end{itemize}
The proposed \mname \ achieves state-of-the-art performance on the structure-based drug design (SBDD), demonstrating superior binding affinities and structural validity (e.g., Vina score of -7.21).

\section{Related Work}

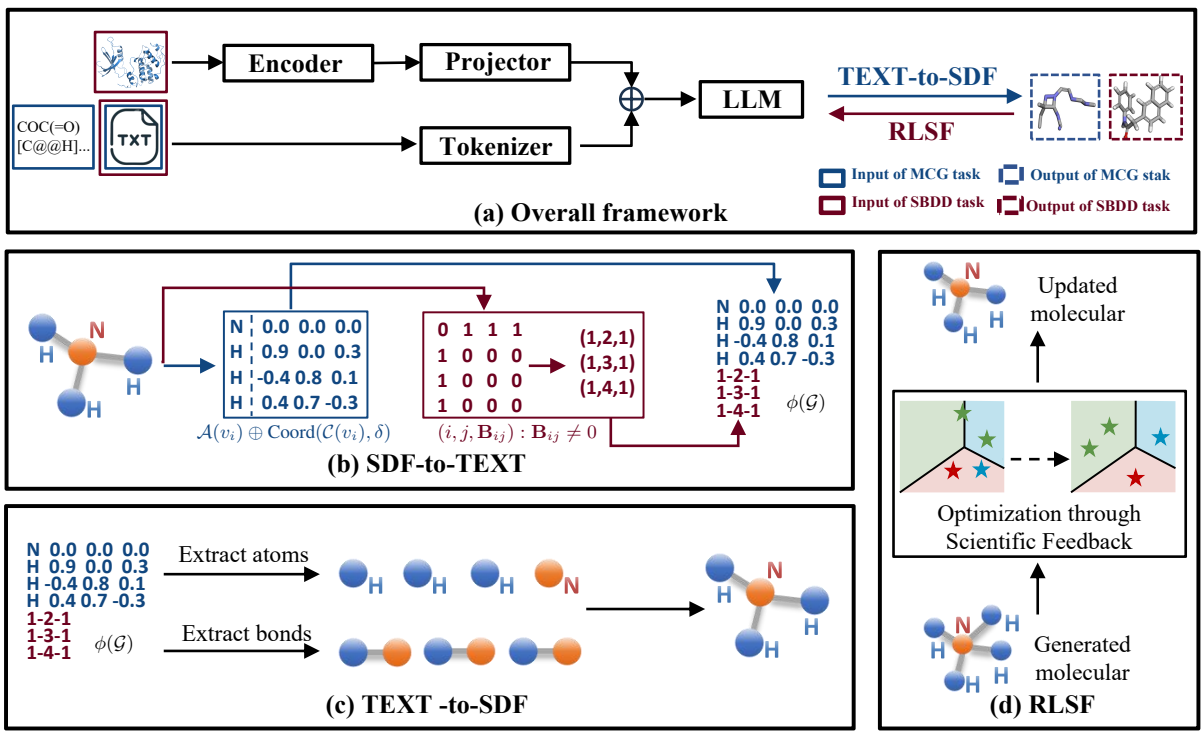
\begin{figure*}[ht]
    \footnotesize\centering
    \begin{overpic}[width=0.9\linewidth]{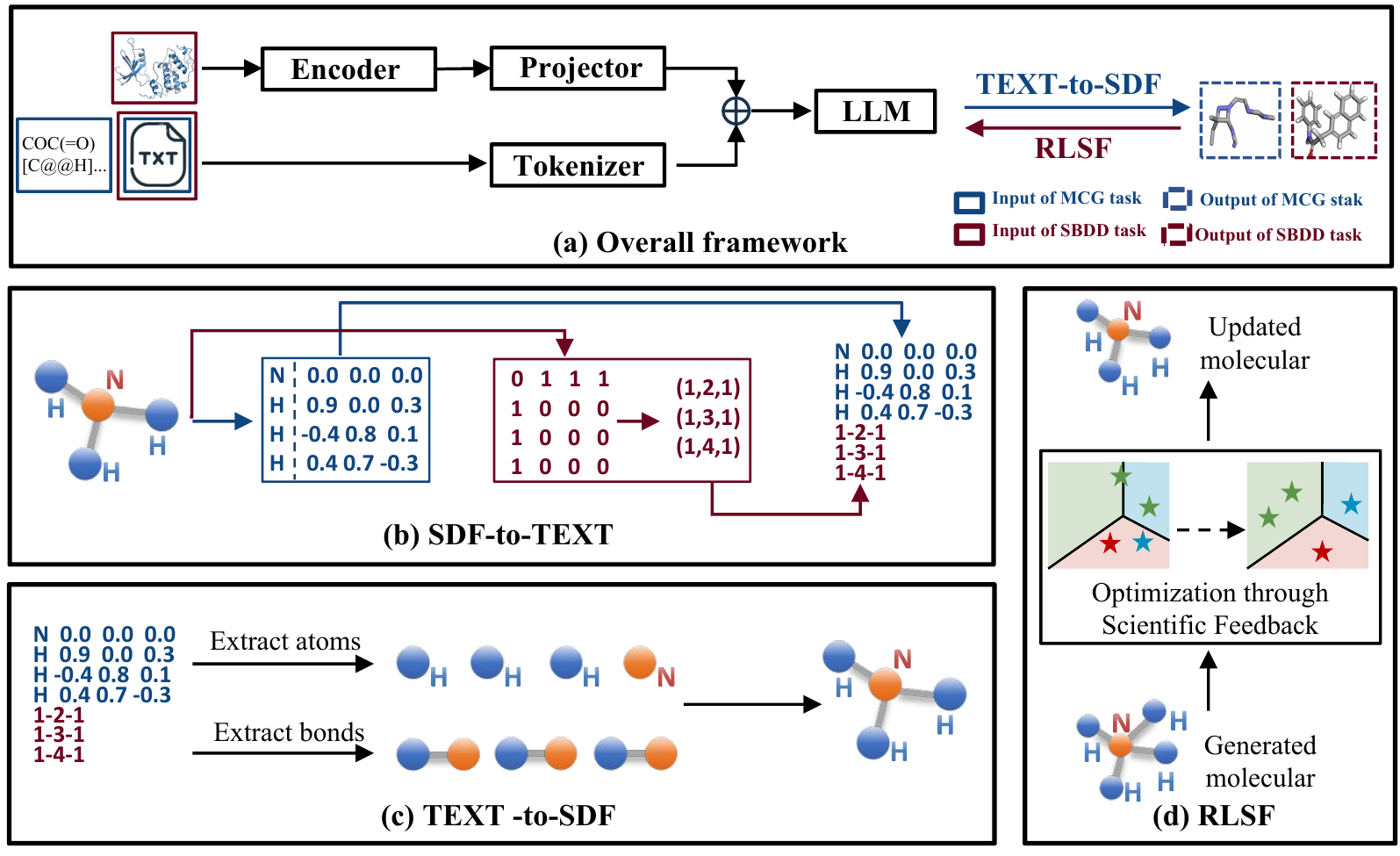}

        \put(16.5,25){\scalebox{0.8}{$\textcolor[rgb]{0.0627,0.2745,0.5019}{\mathcal{A}(v_i) \oplus \text{Coord}(\mathcal{C}(v_i), \delta)}$}}

        \put(36.5,25){\scalebox{0.8}{$\textcolor[rgb]{0.4274,0.0039,0.1215}{ (i,j,\mathbf{B}_{ij}) : \mathbf{B}_{ij} \neq 0}$}}

        \put(65.5,27.5){\scalebox{0.8}{$\phi(\mathcal{G})$}}

        \put(8,7.5){\scalebox{0.8}{$\phi(\mathcal{G})$}}

    \end{overpic}
    \caption{Framework of \mname. (a) Overall framework: The proposed \mname model is capable of simultaneously performing Molecular Conformation Generation and Structure-based Drug Design tasks. (b) SDF-to-TEXT: 3D structures are parsed into atomic coordinates and bonding matrices, then discretized as token sequences. (c) TEXT-to-SDF: token sequences are decoded to reconstruct 3D molecular graphs. (d) RLSF: reinforcement learning refines generation with scientific feedback.}
    \label{fig:framework}
\end{figure*}

\noindent\textbf{Molecular Conformation Generation (MCG)}. 
MCG reconstructs plausible 3D structures from 2D molecular graphs, constituting a fundamental computational chemistry task. While traditional molecular dynamics and force-field methods exhibit computational limitations and restricted conformational diversity, contemporary deep generative approaches learn conformation distributions directly from data~\cite{xu2022geodiff, zhu2022direct, shi2021learning, yu2024tensorvae}. Early two-stage paradigms predict interatomic distances before coordinate reconstruction~\cite{shi2021learning,xu2021end}, while subsequent direct coordinate generation employs equivariant models~\cite{zhu2022direct, yu2024tensorvae}. Diffusion-based frameworks~\cite{xu2022geodiff,wang2025gadiff,xiao2025dihedralsdiff} enhance sample diversity through stochastic process reversal. Alternative methodologies incorporate thermodynamic priors via score matching~\cite{zhang2023sdegen}, reinforcement learning~\cite{fu2022reinforced}, or GFlowNet sampling~\cite{volokhova2024towards,volokhova2025torsional}. However, generating thermodynamically stable conformations remains challenging. 
 Our method addresses this by integrating stability-aware rewards and reinforcement learning within a protein-conditioned LLM framework, complementing existing generative paradigms with principled optimization of binding and feasibility.

\noindent\textbf{Structure-based Drug Design}. 
Structure-based drug design (SBDD) uses 3D protein structures to generate high-affinity ligands \cite{kuntz1992structure}. Traditional docking-based approaches \cite{clearance} suffer from limited coverage and high cost. Recent deep generative models automate ligand design via reinforcement learning \cite{popova2018deep, fu2022reinforced}, 3D CNNs \cite{crossdock}, or geometric priors \cite{isert2023structure}. Advanced 3D frameworks generate molecules within binding sites \cite{luo20213AR, li2021structure}, enhanced by equivariant diffusion models \cite{schneuing2024structure, qu2024molcraft} and fragment generation \cite{fu2024fragment}.
However, challenges remain in aligning molecular topology with 3D geometry and generalizing across binding sites. Our unified multimodal framework addresses these via joint protein-ligand encoding and reward-guided optimization.

\section{Method}

We propose \textbf{Chem3DLLM}, a unified protein-conditioned multimodal framework (see Figure~\ref{fig:framework}(a)). Chem3DLLM jointly addresses molecular conformation generation (MCG) and structure-based drug design via three key innovations: (1) a Reversible Compression of Molecular Tokenization (RCMT) that preserves 3D structural information, (2) a multimodal large language model integrating protein structures and molecular conformations, and (3) a Reinforcement Learning with Scientific Feedback (RLSF) module that guides generation toward chemically valid structures.
Our approach operates through a three-stage pipeline where molecular structures are first converted into compressed textual representations, then processed by a multimodal LLM conditioned on protein pocket embeddings, and finally optimized using reinforcement learning with chemically-informed rewards.
All the mathematical notations and their explanation are summarized in \textbf{Appendix A} for clarity and consistency.

\subsection{RCMT: Reversible Compression of Molecular Tokenization}

{As discussed in \textbf{Challenge 1} of Section~\ref{sec_Introduction}, a central obstacle in structure-aware molecular generation lies in representing complex 3D molecular conformation in a discrete, learnable format suitable for language models. Our goal is to design a Reversible Compression of 3D Molecular Tokenization that is both compact and structurally complete, enabling efficient integration with LLMs while preserving critical geometric and chemical features. Specifically, we develop a mathematically rigorous bidirectional encoding scheme that transforms SDF molecular files into compressed textual sequences with guaranteed lossless reconstruction. This representation supports efficient storage, reversible processing, and seamless integration into language-based generative frameworks.} 
Please refer to Figure~\ref{fig:framework} (b) and (c) for a toy example of the compression method.

\noindent
\textbf{{Molecular Graph Formalization.}} We formally represent a molecule as a labeled graph $\mathcal{G} = (\mathcal{V}, \mathcal{E}, \mathcal{A}, \mathcal{C})$, where $\mathcal{V} = \{v_1, v_2, \ldots, v_N\}$ denotes the set of $N$ atoms, $\mathcal{E} \subseteq \mathcal{V} \times \mathcal{V}$ represents the bond connectivity, $\mathcal{A}: \mathcal{V} \rightarrow \Sigma$ maps atoms to their chemical symbols from alphabet $\Sigma$, and $\mathcal{C}: \mathcal{V} \rightarrow \mathbb{R}^3$ assigns 3D coordinates to each atom.

\noindent
\textbf{{Encoding Algorithm.}} 
\textcolor{black}{Given $\mathcal{G}$, we define an encoding function $\phi: \mathcal{G} \rightarrow \mathcal{T}$ that generates a compact sequence $\mathcal{T} = \{t_1, t_2, \ldots, t_k\}$ encoding atomic and bond-level information. For each atom $v_i$, we concatenate its chemical symbol with a quantized version of its 3D coordinate:}
\begin{align}
& \alpha_i = \mathcal{A}(v_i) \oplus \text{Coord}(\mathcal{C}(v_i), \delta), \\
& \text{Coord}(\mathbf{r}, \delta) = \lfloor \mathbf{r} / \delta \rfloor \cdot \delta, \; \delta = 10^{-4}.
\end{align}
\textcolor{black}{Here, $\delta$ controls quantization granularity and ensures numerical stability while maintaining sub-angstrom precision.}

\textcolor{black}{To represent chemical bonds, we construct an adjacency matrix $\mathbf{B} \in \{0,1,2,3,4\}^{N \times N}$, where $\mathbf{B}_{ij}$ specifies the bond order between atom $i$ and atom $j$:}
\begin{equation}
\mathbf{B}_{ij} = \begin{cases}
1 & \text{if } (v_i, v_j) \in \mathcal{E} \text{ and bond type is single}, \\
2 & \text{if } (v_i, v_j) \in \mathcal{E} \text{ and bond type is double}, \\
3 & \text{if } (v_i, v_j) \in \mathcal{E} \text{ and bond type is triple}, \\
4 & \text{if } (v_i, v_j) \in \mathcal{E} \text{ and bond type is aromatic}, \\
0 & \text{otherwise}.
\end{cases}
\end{equation}
To achieve compression, we apply run-length encoding (RLE)~\citep{robinson2005results} (a well-known compression scheme) to the sparse bond matrix:
\begin{align}
\phi(\mathcal{G}) = \text{Concat}(\{\alpha_i\}_{i=1}^N, \{(i,j,\mathbf{B}_{ij}) : \mathbf{B}_{ij} \neq 0\}). 
\end{align}

\noindent
\textbf{{Decoding Algorithm.}} The inverse function $\phi^{-1}: \mathcal{T} \rightarrow \mathcal{G}$ reconstructs the molecular graph by parsing the text sequence and extracting atomic coordinates, symbols, and bond connectivity:
\begin{equation}
\phi^{-1}(\mathcal{T}) = \mathcal{G}(\text{ExtractCoord}(\alpha_i), \text{ExtractSymbol}(\alpha_i), \{b_{ij}\}). 
\end{equation}

\noindent
\textbf{{Compression Analysis.}} The compression ratio $\rho$ is theoretically bounded by:
\begin{equation}
\begin{aligned}
\rho & = \frac{|\text{SDF}(\mathcal{G})|}{|\phi(\mathcal{G})|} \\ 
& \geq \frac{3N \cdot \log_{10}(L) + N \cdot |\Sigma|}{N \cdot (|\Sigma| + 3 \cdot \log_{10}(\delta^{-1})) + |\mathcal{E}| \cdot \log_{10}(4)}, 
\end{aligned}
\end{equation}
where $L$ represents the typical coordinate range and $|\Sigma| = 118$ is the size of the periodic table $\Sigma$. 

\noindent
\textbf{{Lossless Reconstruction.}} 
\textcolor{black}{Our encoding scheme ensures complete preservation of molecular geometry and connectivity, satisfying $\mathcal{A}(v_i) = \mathcal{A}'(v_i)$ and $\mathcal{E} = \mathcal{E}'$ after decoding. Across the QM9 dataset, our method achieves an average compression ratio $\rho > 3.2$ (an empirical validation in Figure~\ref{fig:RCMT_Components_Analysis}) with 98.56\% molecular validity and RMSD = 0, demonstrating the effectiveness of the reversible representation. The algorithm operates in linear time and space $\mathcal{O}(N + |\mathcal{E}|)$, making it scalable to large molecular libraries.} 

\subsection{Multimodal LLM Architecture}

As discussed in \textbf{Challenge 2} of Section~\ref{sec_Introduction}, effectively integrating protein context into molecule generation remains a critical barrier for structure-aware drug design. To address this fundamental limitation, we propose a multimodal LLM architecture inspired by LLaVA~\cite{liu2023llava} that seamlessly bridges protein structural information with molecular generation capabilities.

\noindent
\textbf{{Architecture Overview.}} Our multimodal LLM consists of three key components as shown in Figure~\ref{fig:framework} (a): (1) a protein encoder $\mathcal{E}_{prot}$ based on Evolutionary Scale Modeling (ESM)~\cite{hsu2022ESM} (a pretrained protein language model), (2) a language model $\mathcal{L}_{mol}$ Qwen2-7B~\cite{team2024qwen2}, and (3) a cross-modal alignment module $\mathcal{A}_{align}$ that bridges protein and textual representations. 

\noindent
\textbf{{Protein-Molecule Integration.}} To integrate protein structural information with molecular generation, we employ a cross-modal alignment framework. Given a protein pocket $\mathcal{P} = \{r_1, r_2, \ldots, r_n\}$ containing $n$ residues, we extract contextualized embeddings using the pre-trained ESM model $\mathcal{E}_{prot}(\mathcal{P})$. The protein embedding is subsequently projected into the language model's hidden space through a multi-layer perceptron: $\mathbf{h}_{aligned} = \text{MLP}(\mathbf{h}_{pocket})$, where the MLP consists of two linear layers with GELU activation to align with the language model's hidden dimension.

\noindent
\textbf{{Multimodal Generation.}} The aligned protein representation is concatenated with text embeddings and fed into the language model. Given a textual prompt $\mathbf{T} = \{t_1, t_2, \ldots, t_m\}$ and protein context $\mathbf{h}_{aligned}$, our model generates molecular representations autoregressively:
\begin{equation}
P(\mathcal{M}|\mathbf{T}, \mathcal{P}) = \prod_{i=1}^{|\mathcal{M}|} P(m_i | m_{<i}, [\mathbf{T}; \mathbf{h}_{aligned}]; \theta), 
\end{equation}
where $\mathcal{M}$ denotes the target molecular representation in our proposed text format, $[\cdot; \cdot]$ represents concatenation, and $\theta$ are the model parameters.

\noindent
\textbf{{Training Strategy.}} We adopt a two-stage training approach to optimize our multimodal framework: (1) {Supervised Fine-tuning (SFT)}: We train the model on instruction-following data for protein-conditioned molecular generation tasks. During this stage, we freeze the pre-trained ESM encoder and only optimize the projection module (MLP) and the language model backbone. (2) {Reinforcement Learning (RL)}: We further optimize the model using reinforcement learning to improve chemical validity and molecular properties, as detailed in the following subsection.

The SFT training objective uses the standard next-token prediction loss:
\begin{equation}
\mathcal{L}_{SFT} = -\sum_{t=1}^{T} \log P(y_t | y_{<t}, [\mathbf{x}_{text}; \mathbf{h}_{aligned}]),
\end{equation}
where $y_t$ represents the target token at position $t$, $\mathbf{x}_{text}$ denotes the input text instruction tokens, $\mathbf{h}_{aligned}$ is the aligned protein representation, and $[\cdot; \cdot]$ represents concatenation of the text and protein inputs.

\subsection{Reinforcement Learning with Scientific Feedback (RLSF)}
While our multimodal LLM provides a strong prior for conformation generation, it does not inherently guarantee the chemical stability or synthesizability of the generated molecules. As discussed in \textbf{Challenge 3} of Section~\ref{sec_Introduction}, generating thermodynamically plausible and chemically valid structures remains a nontrivial goal. To address this challenge, we introduce a Reinforcement Learning with Scientific Feedback (RLSF) framework that refines the pretrained generative model using chemically grounded reward signals,
as illustrated in Figure~\ref{fig:framework}(d).

\noindent
\textbf{{Reward Function Design.}} We formulate a stability-centric reward function that prioritizes molecular stability ($S_m$) and atomic stability ($S_a$) over conventional drug-likeness metrics:

\begin{equation}
R(m) = \alpha \cdot S_m(m) + \beta \cdot S_a(m) + \gamma \cdot D(m) + \delta \cdot V(m),
\end{equation}
where $m$ denotes the generated molecule, $\alpha$ and $\beta$ are the primary weighting parameters for molecular and atomic stability, respectively (with $\alpha + \beta > \gamma + \delta$), $S_m(m)$ represents molecular stability, $S_a(m)$ denotes atomic stability, $D(m)$ is the molecular diversity metric and $V(m)$ is the chemical validity indicator.

The molecular stability $S_m(m)$ is calculated based on the overall molecular energy:
\begin{equation}
S_m(m) = \exp\big(-\frac{E_{total}(m) - E_{ref}}{k_B T}\big). 
\end{equation}
where $k_B$ is the Boltzmann constant, and $T$ is the temperature. $E_{total}(m)$ denotes the computed total energy of the generated molecule $m$, and $E_{ref}$ is a reference energy value derived from stable molecules in the training corpus, respectively. 
The atomic stability $S_a(m)$ evaluates the local chemical environment stability for each atom:
\begin{equation}
S_a(m) = \frac{1}{N} \sum_{i=1}^{N} \exp\big(-\frac{E_{local}(a_i) - E_{local,ref}(a_i)}{k_B T}\big),
\end{equation}
where $N$ is the number of atoms, $E_{local}(a_i)$ is the local energy of atom $i$ within the molecular context, $E_{local,ref}(a_i)$ is the reference local energy for atom type $i$.

\noindent
\textbf{{RLSF Optimization.}} 
Utilizing a pretrained model as the initial policy $\pi_{\theta_0}$, we employ Proximal Policy Optimization (PPO)~\citep{schulman2017ppo} to maximize the stability-driven expected reward:

\begin{equation}
\begin{aligned}
L^{CLIP}(\theta) = \mathbb{E}_t \Big[ &\min\big( r_t(\theta) \hat{A}_t, \\
&\text{clip}(r_t(\theta), 1-\epsilon, 1+\epsilon) \hat{A}_t \big) \Big],
\end{aligned}
\end{equation}
where $r_t(\theta) = \frac{\pi_\theta(a_t|s_t)}{\pi_{\theta_{old}}(a_t|s_t)}$ is the probability ratio, $\hat{A}_t$ is the advantage estimate, and $\epsilon$ is the clipping parameter. 
The policy gradient update follows:
\begin{equation}
\theta_{k+1} = \theta_k + \alpha_{lr} \nabla_\theta L^{CLIP}(\theta_k), 
\end{equation}
where $\alpha_{lr}$ denotes learning rate. 
This RLSF approach prioritizes molecular and atomic-level stability to guide the model toward generating chemically stable and thermodynamically feasible molecular structures. The stability-oriented reward function ensures that generated molecules exhibit reasonable thermodynamic properties and maintain proper atomic coordination environments, thereby enhancing their practical applicability in chemical synthesis and drug discovery.

\section{Experiments}

\subsection{Tasks and Datasets}

\noindent\textbf{{Task 1: Molecular Conformation Generation (MCG).}}
Given a SMILES input, the model generates the corresponding 3D conformation. We evaluate on the QM9 dataset~\citep{ramakrishnan2014quantum}, which contains 130K small molecules with 3D structures and quantum properties (up to 29 atoms including hydrogens). Following~\citet{anderson2019cormorant}, we adopt the standard 100K/18K/13K train/val/test split and generate 10,000 molecules for evaluation.

\noindent\textbf{{Task 2: Structure-Based Drug Design (SBDD).}} 
Given a protein binding pocket, the model generates 3D molecules with potential binding affinity. We evaluate on the CrossDocked dataset~\citep{crossdock}, containing 100K training protein-ligand pairs and 100 test proteins, with 100 generated molecules per test protein.

\begin{table*}[!h]
\centering


\begin{tabular}{lccccc}
\toprule[0.8pt]
\textbf{Method} & \textbf{Atom Stability (\%)} & \textbf{Mol Stability (\%)} & \textbf{Valid (\%)} & \textbf{Unique (\%)}  \\ 
\hline
ENF & 85.0 & 4.9 & 40.2 & 39.4  \\
G-SchNet & 95.7 & 68.1 & 85.5 & 80.3 \\
GDM & 97.0 & 63.2 & - & - \\
GDM-AUG & 97.6 & 71.6 & 90.4 & 89.5  \\
EDM & 98.7 & 82.0 & 91.9 & 90.7 \\
EDM-Bridge & 98.8 & 84.6 & 92.0 & 90.7  \\
GeoLDM & 98.9  & 89.4  & 93.8  & 92.7   \\
RDKit & 98.25 & 86.87 & 100.0 & 100.0  \\
\hline

\textbf{Chem3DLLM (Ours)} & \textbf{99.45} & \textbf{95.00} & \textbf{100.00} & \textbf{100.00}  \\
\bottomrule[0.8pt]
\end{tabular}
\caption{Results on molecular conformation generation task. }
\label{tab:conformation}
\end{table*}

\subsection{Experimental Setup}

\noindent\textbf{Implementation Details.} Our framework is implemented using PyTorch and trained on 8 NVIDIA A800 GPUs, each with 80 GB of memory. 
We adopt Qwen2-7B as the base language model.
The optimization is performed using the Adam optimizer with a learning rate of $1\times10^{-4}$ and a batch size of 32. 
The protein structure projection module consists of 2 MLP layers with 3584 hidden dimensions. 
For RLSF optimization, we employ PPO with standard hyperparameters: a clipping parameter of $\epsilon = 0.2$ and discount factor $\gamma = 0.99$.

\noindent\textbf{Training Protocol.} We adopt a two-stage training strategy: (1) SFT on the Cross-Docked dataset for 3 epochs, and (2) fine-tuning on QM9 with reinforcement learning for 150 steps using our stability-based reward function.

\noindent\textbf{Baseline Methods.} 
For MCG task, we compare with early equivariant generative models, including G-SchNet\citep{gebauer2019symmetry} and Equivariant Normalizing Flows\citep{garcia2021n}, as well as recent diffusion-based methods such as EDM, EDM-Bridge~\citep{wu2022diffusion}, GDM~\citep{hoogeboom2022equivariant}, and GeoLDM~\citep{xu2023geoldm}. RDKit~\citep{landrum2006rdkit}, an open-source cheminformatics toolkit, is also included as a classical non-learning baseline. 
Comparison results are adapted from GeoLDM.
For the SBDD task, we compare with autoregressive models, including AR~\citep{luo20213AR}, Pocket2Mol~\citep{peng2022pocket2mol}, and FLAG~\citep{zhang2023learning}, as well as diffusion-based methods TargetDiff~\citep{guan20233d} and DecompDiff~\citep{guan2024decompdiff}, which includes Decomp-R (with reference priors), Decomp-O (with AlphaSpace2~\citep{katigbak2020alphaspace}) and MolCRAFT~\citep{qu2024molcraft}. 
Comparison results are adapted from MolCRAFT. \textbf{Appendix B} provides a comprehensive description of the baseline methods used for comparison.

\noindent\textbf{Evaluation Metrics}. 
For the MCG task, we use comprehensive metrics to assess both chemical validity and structural accuracy:
(1) \textbf{Atom Stability (\%):} Percentage of atoms with stable local chemical environments;
(2) \textbf{Molecular Stability (\%):} Percentage of molecules with overall thermodynamic stability; 
(3) \textbf{Validity (\%):} Percentage of generated molecules that satisfy chemical rules; 
(4) \textbf{Uniqueness (\%):} Percentage of unique molecules in generated samples. 
For structure-based drug design evaluation, we primarily use  \textbf{Vina Score} to assess binding affinity between target proteins and generated ligands via AutoDock Vina~\cite{trott2010autodock}. We report both the direct Vina Score of the generated pose and Vina Min., which scores the energy-minimized optimized pose.

\subsection{Main Results and Analysis}

To comprehensively assess the effectiveness and versatility of Chem3DLLM, we conduct evaluations from two complementary perspectives: task generalization and joint-task optimization. The first focuses on the model's ability to adapt to different molecular generation tasks under single-task supervision, while the second investigates its performance under unified multi-task training.

\noindent\textbf{Generalization Across Molecular Tasks.} We begin by evaluating Chem3DLLM under single-task training on each task independently. Despite utilizing the same architecture and tokenization strategy, Chem3DLLM achieves state-of-the-art performance on both tasks. Table~\ref{tab:conformation} summarizes the performance on the conformation generation task. 
Chem3DLLM surpasses all previous baselines, including GeoLDM, and even outperforms RDKit which relies on distance geometry and force-field-based optimization in all metrics. 
Additionally, our model attains \textbf{95.00\%} molecular stability, \textbf{100.00\%} chemical validity, and \textbf{100.00\%} uniqueness, highlighting the effectiveness of our reversible tokenization and stability-oriented reward design in preserving 3D chemical fidelity.

Table~\ref{tab:sbdd} reports SBDD results, where Chem3DLLM achieves a best median docking score of \textbf{-7.15} and an average of \textbf{-7.03}, outperforming all baselines. Compared to MolCRAFT (-6.59 average), our model improves by 0.42 on average and over 5 in the best case. We argue that these gains arise from the protein-aware decoder’s ability to capture long-range protein-ligand interactions, as well as the pretrained molecular embedding space, which enforces geometric plausibility and biological compatibility.

\begin{table}[h]
\centering
\begin{tabular}{lcc}
\toprule[0.9pt]
\textbf{Methods} & \textbf{Vina Score Avg.(↓) } & \textbf{Vina Med.(↓)} \\
\hline
Reference & -6.36 & -6.46 \\
AR & -5.75 & -5.64 \\
Pocket2Mol & -5.14 & -4.70 \\
FLAG$^\dagger$ & 16.48 & 4.53 \\
TargetDiff & -5.47 & -6.30 \\
Decomp-R & -5.19 & -5.27 \\
Decomp-O & -5.67 & -6.04 \\
MolCRAFT & -6.59 & -7.04 \\
\hline
\textbf{Chem3DLLM} & \textbf{-7.03} & \textbf{-7.15}\\
\bottomrule[0.9pt]
\end{tabular}
\caption{Results on structure-based drug design task. ``Med.'' means median.}
\label{tab:sbdd}
\end{table}

\begin{table}[h]
\centering
\begin{tabular}{lcc}
\toprule
\textbf{Method} & \textbf{Vina Score Avg.(↓)} & \textbf{Vina Min.(↓)} \\
\midrule
\textbf{Chem3DLLM} & -7.03 & -12.20 \\
\textbf{Chem3DLLM \textsuperscript{†}} & \textbf{ -7.21} &\textbf{-12.30} \\
\bottomrule
\end{tabular}
\caption{Results of incorporating additional geometric supervision into the SBDD task. \textbf{Chem3DLLM\textsuperscript{†}} represents the same model jointly trained on both conformation and SBDD tasks.}
\label{tab:sbdd_full}
\end{table}

\noindent\textbf{Joint Optimization under Multi-Task Supervision.} Structure-based drug design is a core task in molecular discovery with high practical value and complexity. Unlike conformation generation, which focuses on geometric fidelity, SBDD requires modeling multi-modal interactions between protein structures and chemical graphs. Thus, its performance reflects the model’s capacity to reason across molecular and biological modalities, making it an ideal benchmark for assessing joint-task capabilities.

Table~\ref{tab:sbdd_full} shows that when jointly trained with the conformation task, Chem3DLLM achieves further improvements on SBDD, with a best docking score of \textbf{-12.30} and an average score of \textbf{-7.21}. This multi-task setup does not degrade SBDD performance—on the contrary, it enhances it. We attribute this gain to auxiliary geometric supervision from conformation prediction, which injects spatial inductive bias into ligand generation and strengthens the chemical validity of synthesized candidates. Notably, the performance improvement from \textbf{-7.03} to \textbf{-7.21} demonstrates that multi-task learning yields a synergistic effect rather than interference. This result highlights Chem3DLLM's capability to support both molecular and structural reasoning within a unified architecture, without requiring task-specific compromises.

\begin{figure}[ht]
\centering
\includegraphics[width=\columnwidth]{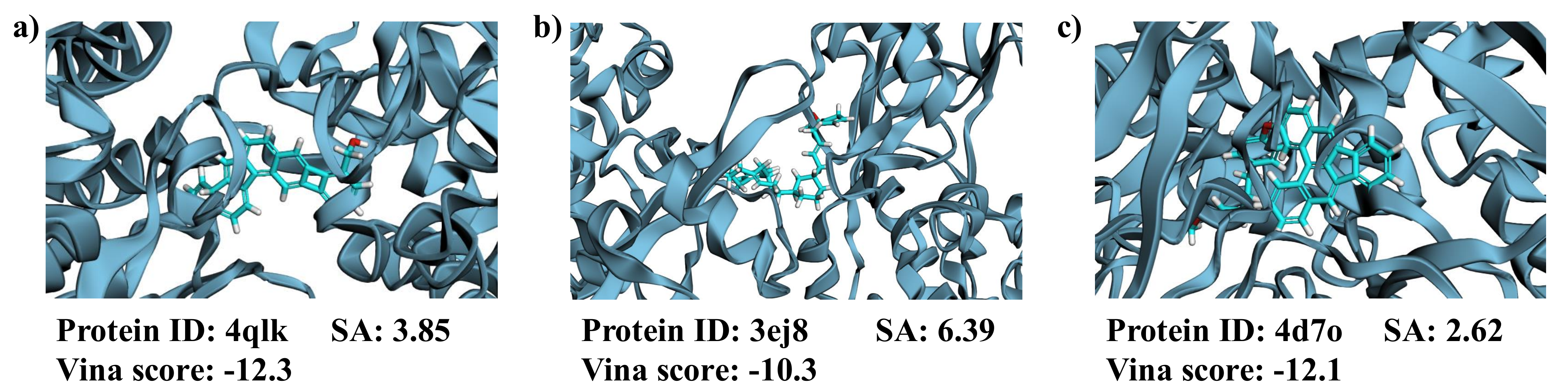} 
\caption{Case studies on structure-based drug design task. }
\label{fig:case2}
\end{figure}
\begin{figure}[h!]
\centering
\includegraphics[width=\columnwidth]{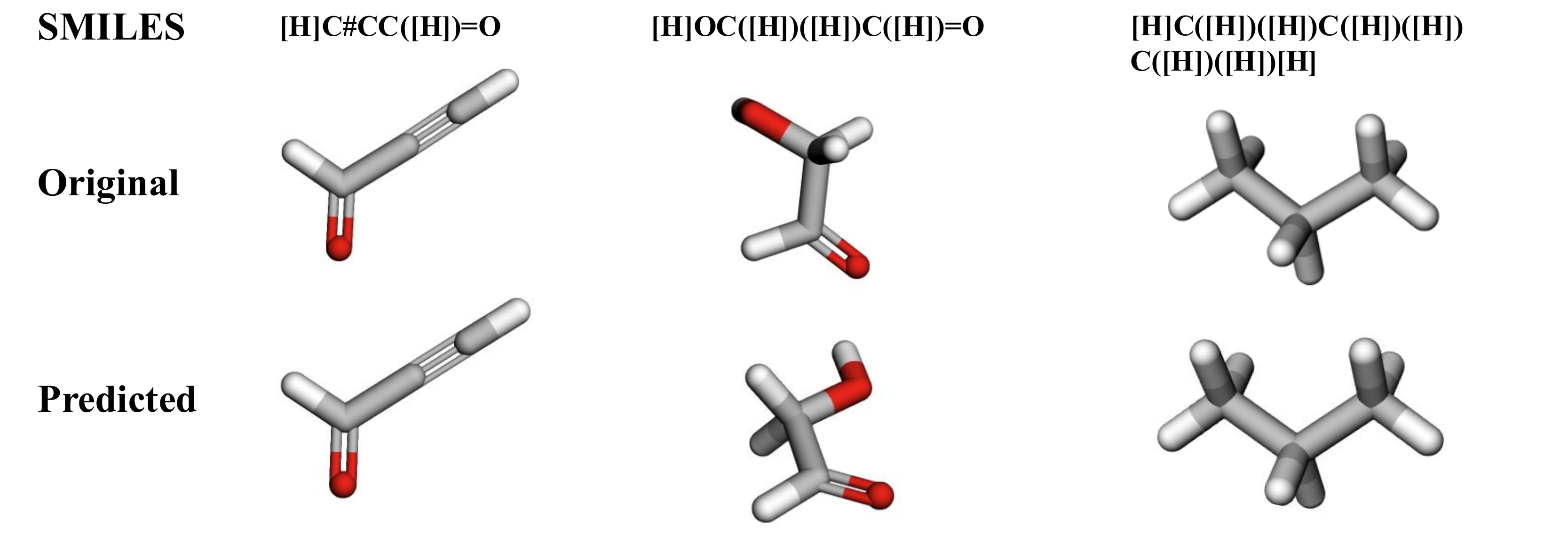} 
\caption{Case studies on molecular conformation generation task. }
\label{fig:case1}
\end{figure}

\noindent\textbf{Qualitative Results.} We further present qualitative results to visually assess the performance of our model on both MCG and SBDD tasks. As shown in Figure~\ref{fig:case2}, on SBDD tasks, the generated molecules bind tightly with protein pockets, with Vina scores below $-10$ and favorable 
synthetic accessibility (SA) metrics.
The structures exhibit a strong balance between synthetic accessibility and pharmacological relevance, demonstrating \mname’s ability to generate bioactive, chemically plausible ligands across diverse targets.

To assess 3D geometry reconstruction, Figure~\ref{fig:case1} compares predicted conformations with ground truth. Given SMILES input, Chem3DLLM generates chemically valid 3D structures with high fidelity across diverse motifs (e.g., aliphatic chains, carbonyls, triple bonds). The close visual alignment confirms the ability of our proposed Chem3DLLM to capture fine-grained atomic arrangements and stereochemistry, qualitatively supporting its structural accuracy and generality. \textbf{Appendix C} presents additional visualizations for conformation generation and SBDD tasks, and \textbf{Appendix D} provides visual examples of molecular compression from SDF files to tokenized sequences.

\subsection{Ablation Studies}

\begin{figure}[h!]
\centering
\includegraphics[width=\columnwidth]{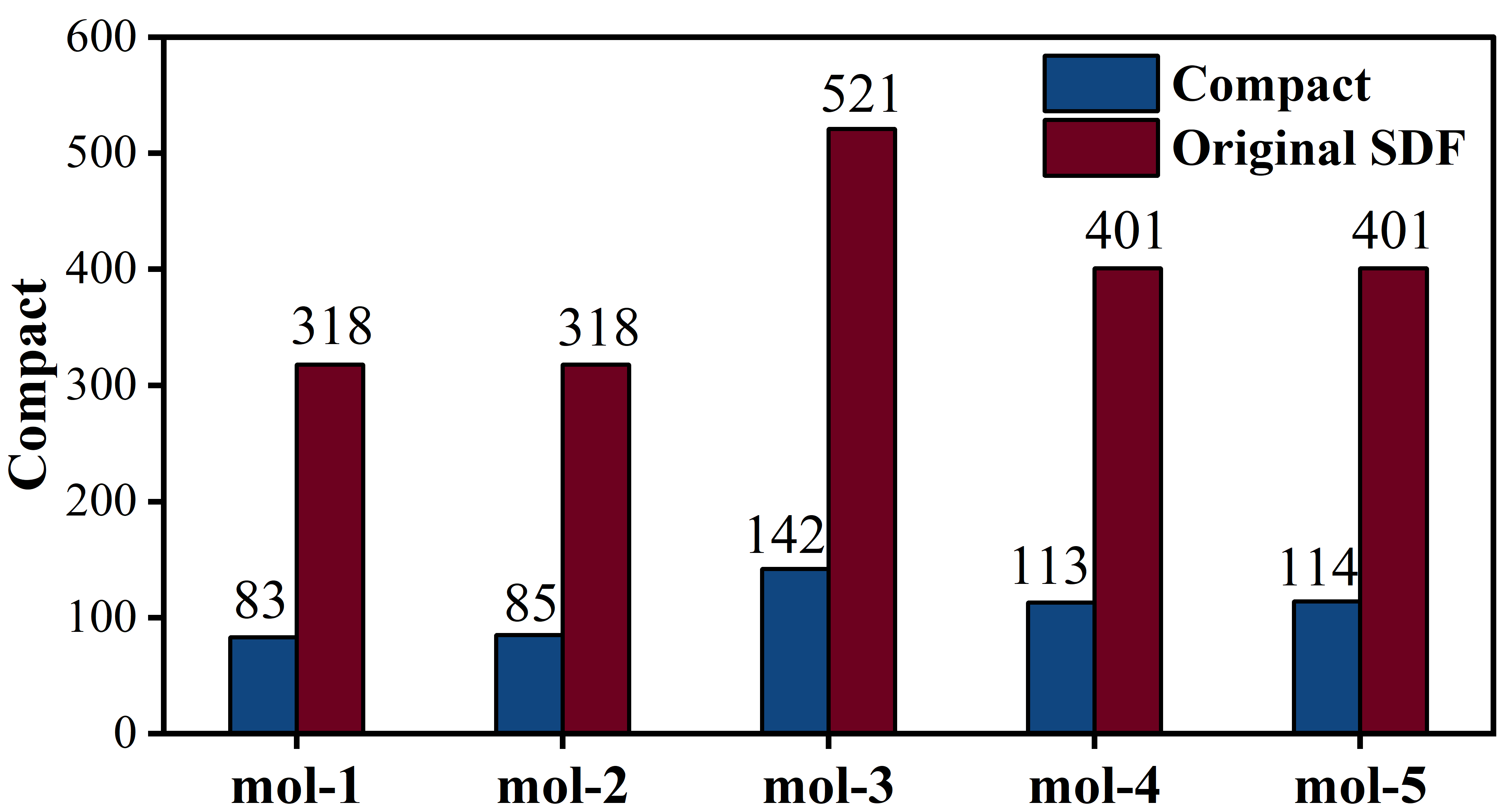} 
\caption{Visualization of character counts for five representative molecules under two encoding schemes: original SDF and our proposed RCMT-based compact textual format. 
}
\label{fig:RCMT_Components_Analysis}
\end{figure}

\noindent\textbf{Reversible Compression Analysis.} To support efficient molecular representation within large language models, our RCMT framework is designed to produce compact, information-preserving token sequences. As illustrated in Figure~\ref{fig:RCMT_Components_Analysis}, we compare the character count of the original SDF format against the compact representation produced by RCMT across five example molecules. The results demonstrate a significant reduction in token length, with compression rates exceeding 60\% in most cases. On average, across a randomly selected set of 100 molecules, our method achieves a character compression rate of 35.20\%. This substantial reduction highlights the advantage of RCMT in reducing memory and computational overhead during downstream generative modeling, without compromising on molecular fidelity or 3D structure precision.

\begin{table}[h]
\centering


\begin{tabular}{lcc}
\toprule
\textbf{Method} & \textbf{Vina Score Avg.(↓)} & \textbf{Vina Min(↓)} \\
\midrule
w/o RCMT & -1.82 &-4.70\\
w/o RLSF (SFT only) & -7.03 &-12.20\\
\textbf{Chem3DLLM} & \textbf{ -7.21} &\textbf{-12.30} \\
\bottomrule
\end{tabular}
\caption{Ablation study on key components and training strategies.}
\label{tab:ablation_components}
\end{table}

\noindent\textbf{Architectural and Training Contributions.} We perform ablation studies to isolate the effects of the representation scheme (RCMT) and the training objective (reinforcement learning). As shown in Table~\ref{tab:ablation_components}, removing RCMT leads to the weakest performance (Vina score: -1.82), highlighting limited structural awareness. Incorporating RCMT alone (w/o RLSF) improves the score to -7.03, demonstrating its benefit for conformation learning. The full Chem3DLLM model, with both RCMT and RLSF, achieves -7.21. These results indicate that while RLSF offers task-specific adaptation, RCMT plays the primary role by enabling expressive, compact encodings compatible with language modeling.

\section{Conclusion}
\label{sec_Conclusion}

In this paper, we have proposed \mname, a unified, protein-conditioned framework that bridges the long-standing gap between 3-D molecular structure fidelity and chemical validity in structure-based drug design.  
By (i) compressing full atomic 3D geometries into a lossless, sequence-compatible representation, (ii) aligning heterogeneous protein–ligand modalities in the token embedding space, and (iii) steering generation with stability-aware RL rewards, \mname\  achieves a Vina score of -7.21 on the CrossDocked benchmark—outperforming recent diffusion- and autoregressive baselines while maintaining 100\% syntactic validity and $>$95\% synthetic accessibility.  
These results demonstrate that explicit handling of 3-D constraints and cross-modal alignment inside a single LLM architecture is both feasible and beneficial.  
Future work will explore scaling to larger chemical spaces and integrating downstream assay feedback for end-to-end lead optimization.

\bibliography{aaai2026}

\clearpage
\onecolumn  

\begin{center}
\LARGE\textbf{Appendix File for ``Chem3DLLM: 3D Multimodal Large Language Models for Chemistry''}
\end{center}

This appendix provides additional materials to support the main paper. In \textbf{Appendix~\ref{appendix:Mathematical_notations}}, we present a complete list of mathematical notations used throughout the method and experiments for ease of reference. \textbf{Appendix~\ref{appendix:compared_methods}} details all baseline methods used in our evaluation, including autoregressive and diffusion-based models for both molecular conformation generation (MCG) and structure-based drug design (SBDD). \textbf{Appendix~\ref{appendix:Additional_Visualization_Examples}} complements the qualitative analysis with additional visualization examples for both tasks. Finally, \textbf{Appendix~\ref{appendix:compression_case}} illustrates a case study of the proposed reversible compression scheme (RCMT), showing the transformation from raw SDF format to compact token sequences.

\appendix
\section{Mathematical notations}
\label{appendix:Mathematical_notations}

To facilitate consistent reading and clarify the notation used throughout the paper, we summarize all mathematical symbols and their corresponding definitions in Table~\ref{table:notation_complete}. The listed notations span molecular representations, model components, and training objectives, and are aligned with the terminology used in the main method and experiment sections. This reference table serves as a unified index of all mathematical symbols used throughout the main paper.

\begin{longtable}{c|p{9.99cm}}
\caption{Mathematical notations.} \label{table:notation_complete} \\
\toprule[1pt]
\textbf{Notations} & \textbf{Descriptions} \\
\midrule
\endfirsthead

\multicolumn{2}{c}%
{{\bfseries \tablename\ \thetable{} -- continued from previous page}} \\
\toprule[1pt]
\textbf{Notations} & \textbf{Descriptions} \\
\midrule
\endhead

\midrule \multicolumn{2}{r}{{Continued on next page}} \\
\endfoot

\bottomrule[1pt]
\endlastfoot

$\mathcal{G} = (\mathcal{V}, \mathcal{E}, \mathcal{A}, \mathcal{C})$ & Molecular graph representation: tuple of atoms, bonds, atom symbols, and coordinates \\ 
$N$ & Number of atoms in a molecule \\ 
$\mathcal{V} = \{v_{1}, v_{2}, \ldots, v_{N}\}$ & Set of atoms \\ 
$\mathcal{E} \subseteq \mathcal{V} \times \mathcal{V}$ & Bond connectivity (edge set) \\ 
$\Sigma$ & Alphabet of chemical symbols (periodic table) \\ 
$\mathcal{A}: \mathcal{V} \rightarrow \Sigma$ & Mapping of atoms to chemical symbols \\ 
$\mathcal{C}: \mathcal{V} \rightarrow \mathbb{R}^{3}$ & Mapping of atoms to 3D Cartesian coordinates \\ 
$\phi: \mathcal{G} \rightarrow \mathcal{T}$ & Encoding function from molecular graph to token sequence \\ 
$\mathcal{T} = \{t_{1}, t_{2}, \ldots, t_{k}\}$ & Compressed token sequence representing molecular structure \\ 
$\alpha_{i} = \mathcal{A}(v_{i}) \oplus \textsf{Coord}(\mathcal{C}(v_{i}), \delta)$ & Atom token: chemical symbol concatenated with quantized coordinates \\ 
$\delta$ & Quantization granularity for coordinate discretization \\ 
$\mathbf{r} \in \mathbb{R}^3$ & 3D coordinate vector of an atom \\ 
$\textsf{Coord}(\mathbf{r}, \delta) = [\mathbf{r}/\delta] \cdot \delta$ & Coordinate quantization function \\ 
$\mathbf{B} \in \{0,1,2,3,4\}^{N \times N}$ & Bond adjacency matrix ($0$: no bond; $1$: single; $2$: double; $3$: triple; $4$: aromatic) \\ 
$\phi^{-1}: \mathcal{T} \rightarrow \mathcal{G}$ & Decoding function from token sequence to molecular graph \\ 
$\rho$ & Compression ratio (SDF file size vs. token sequence size) \\ 
$L$ & Typical coordinate range in SDF files \\ 
$|\Sigma|$ & Size of the periodic table (alphabet) \\ 
$\mathcal{E}_{prot}$ & Protein encoder (based on ESM model) \\ 
$\mathcal{L}_{mol}$ & Molecular language model backbone (Quen2-7B) \\ 
$\mathcal{A}_{align}$ & Cross-modal alignment module \\ 
$\mathcal{P} = \{r_{1}, r_{2}, \ldots, r_{n}\}$ & Protein pocket with $n$ residues \\ 
$\mathbf{h}_{aligned}$ & language model's hidden states \\ 
$\mathbf{h}_{pocket}$ & Protein embedding\\ 
$\text{MLP}$ & Multilayer Perceptron as projector \\ 
$P(\mathcal{M}|\mathbf{T},\mathcal{P})$ & Conditional generation probability of molecular representation $\mathcal{M}$ \\ 
$\mathbf{T} = \{t_{1}, t_{2}, \ldots, t_{m}\}$ & Input text prompt tokens \\ 
$[\cdot;\cdot]$ & Concatenation operator \\ 
$\theta$ & Model parameters \\ 
$\mathcal{L}_{SFT}$ & Supervised fine-tuning loss \\ 
$y_{t}$ & Target token at position $t$ \\ 
$\mathbf{x}_{text}$ & Input text instruction tokens \\ 
$R(m)$ & Reward function for generated molecule $m$ \\ 
$\alpha, \beta, \gamma, \delta$ & Weighting parameters in reward function \\ 
$S_{m}(m)$ & Molecular stability score \\ 
$S_{a}(m)$ & Atomic stability score \\ 
$D(m)$ & Molecular diversity metric \\ 
$V(m)$ & Chemical validity indicator \\ 
$E_{total}(m)$ & Computed total energy of molecule $m$ \\ 
$E_{ref}$ & Reference energy from stable molecules \\ 
$k_{B}$ & Boltzmann constant \\ 
$T$ & Temperature \\ 
$E_{local}(a_{i})$ & Local energy of atom $i$ \\ 
$E_{local,ref}(a_{i})$ & Reference local energy for atom type $i$ \\ 
$L^{CLIP}(\theta)$ & PPO clipped surrogate objective \\ 
$r_{t}(\theta)$ & Probability ratio in PPO \\ 
$\hat{A}_{t}$ & Advantage estimate \\ 
$\epsilon$ & Clipping parameter in PPO ($\epsilon=0.2$) \\ 
$\gamma$ & Discount factor in RL ($\gamma=0.99$) \\ 
$\alpha_{lr}$ & Learning rate \\ 

\end{longtable}

\section{Compared Methods}
\label{appendix:compared_methods}

We provide a detailed description of the baseline methods used for comparison in the molecular conformation generation (MCG) and structure-based drug design (SBDD) tasks. These methods are grouped by modeling paradigm (e.g., autoregressive, diffusion-based) and task setting. 

\subsection{Molecular Conformation Generation (MCG) Baselines}

\textbf{-Rule-Based Baseline}

\textbf{RDKit}~\cite{landrum2006rdkit}: RDKit is a widely adopted open-source cheminformatics library that supports molecular modeling, descriptor calculation, and 2D/3D structure manipulation. In our comparison, RDKit serves as a rule-based baseline for molecular conformation generation, where conformers are constructed using distance geometry and force field refinement (e.g., UFF/ETKDG). Although RDKit does not employ data-driven learning, it provides chemically plausible structures with high validity and serves as a classical non-neural benchmark for evaluating generative models.

\textbf{-Equivariant Generative Models}

\textbf{G-SchNet}~\cite{gebauer2019symmetry} is an autoregressive generative model designed to directly synthesize 3D molecular structures in a rotation-equivariant manner. Unlike conventional graph-based or SMILES-based approaches, G-SchNet places atoms sequentially in 3D space by modeling the conditional probability distributions over atom types and distances to previously placed atoms. It incorporates symmetry-aware architectural design via continuous-filter convolutions and auxiliary tokens (focus and origin points) to ensure scalability, local symmetry, and geometric accuracy. Trained on QM9, the model demonstrates high-quality 3D geometry generation and supports property-targeted molecular design through fine-tuning on datasets with specific electronic characteristics.

\textbf{Equivariant Normalizing Flow method}~\cite{garcia2021n} introduces a continuous-time normalizing flow that is equivariant to E(n) symmetries (i.e., rotation, translation, and reflection in Euclidean space). The method leverages E(n) Equivariant Graph Neural Networks (EGNNs) to define the dynamics of a neural ODE, enabling the generation of 3D molecular structures with both spatial and categorical features. Compared to prior approaches, ENF supports exact likelihood estimation and exhibits superior expressivity for modeling molecular distributions. The model is benchmarked on QM9 and synthetic datasets, demonstrating improved negative log-likelihood and better structural stability of generated molecules.

\textbf{-Diffusion-Based Generative Models}

\textbf{GDM}~\cite{hoogeboom2022equivariant}: GDM (Graph Diffusion Model) introduces an E(3)-equivariant denoising diffusion process for 3D molecular generation. It jointly models continuous atomic coordinates and discrete atom types using an equivariant graph neural network (EGNN). The model does not rely on atom ordering, and it achieves superior sample quality and stability by learning to denoise from isotropic Gaussian noise across multiple time steps. Compared to previous autoregressive or normalizing flow methods, GDM offers improved scalability and generation efficiency on datasets such as QM9 and GEOM-Drugs. It also supports log-likelihood evaluation due to its principled probabilistic formulation.

\textbf{EDM-Bridge}~\cite{wu2022diffusion} proposes a novel framework that enhances diffusion-based molecule generation by incorporating informative physical and statistical priors into the diffusion process. Instead of modifying model architectures, EDM-Bridge introduces physically-informed bridge processes that steer the diffusion trajectory towards valid molecular conformations using Lyapunov-guided drifts. The method demonstrates strong performance on 3D molecule generation tasks, achieving better molecular stability and uniformity compared to previous diffusion baselines. Its success stems from effectively injecting domain-specific inductive biases during training, which improves both sample quality and physical plausibility.

\textbf{GeoLDM}~\cite{xu2023geoldm} introduces a geometry-aware diffusion model that explicitly incorporates molecular structural inductive biases by conditioning the generative process on bond lengths, angles, and torsions. It employs a denoising score matching objective over both atomic coordinates and interatomic distances, enabling accurate 3D conformation generation while preserving geometric validity. GeoLDM achieves state-of-the-art performance across multiple benchmarks and serves as a strong diffusion-based baseline in molecular conformation tasks.

\subsection{Structure-Based Drug Design (SBDD) Baselines}

\textbf{-Autoregressive Models}

\textbf{AR}~\cite{luo20213AR} introduces an autoregressive 3D generative model for structure-based drug design, where molecules are generated atom-by-atom conditioned on the 3D protein binding site. At each step, the model predicts a spatial distribution of atom types and samples the next atom accordingly. By progressively expanding the molecular structure, it ensures compatibility with the protein target in 3D space. This autoregressive strategy allows the generation of diverse and chemically valid molecules without relying on post-processing or voxelization, providing strong baseline performance for target-specific molecular generation.

\textbf{Pocket2Mol}~\cite{peng2022pocket2mol}: This method proposes an E(3)-equivariant generative model for structure-based drug design, which directly samples 3D molecular structures conditioned on protein pocket geometry. Pocket2Mol integrates geometric vector perceptrons with vector-based neurons to jointly model atom positions, types, and chemical bonds. It avoids the inefficiency of MCMC-based sampling by directly predicting tractable distributions over atomic coordinates. Experimental results show that Pocket2Mol achieves superior performance in terms of binding affinity, drug-likeness, and synthetic accessibility, while producing chemically realistic substructures and diverse molecular candidates.

\textbf{FLAG}~\cite{zhang2023learning} introduces a generalizable structure-based drug design framework named DrugGPS, which leverages biochemical priors to address the out-of-distribution generalization issue in SBDD. Specifically, it constructs a global interaction graph between protein subpocket prototypes and molecular motifs, and employs a hierarchical 3D graph transformer to capture both atom- and residue-level context. DrugGPS generates ligands motif-by-motif, enriched with global prototype-motif interactions, and achieves superior binding affinity and drug-likeness in challenging out-of-distribution settings.

\textbf{-Diffusion-Based Models}

\textbf{TargetDiff}~\cite{guan20233d} introduces a 3D SE(3)-equivariant diffusion model for target-aware molecule generation. Unlike prior autoregressive or voxel-based approaches, TargetDiff performs non-autoregressive generation in continuous 3D space while maintaining equivariance to rotation and translation. It jointly denoises atom coordinates and types using an equivariant graph neural network conditioned on protein pocket atoms, enabling more realistic binding conformations. In addition to generation, TargetDiff supports unsupervised affinity prediction via learned hidden representations, and achieves strong performance on CrossDocked2020 in both molecular geometry and binding metrics.

\textbf{DecompDiff}~\cite{guan2024decompdiff}: This method introduces a SE(3)-equivariant diffusion model for target-aware molecule generation. Unlike voxel-based or autoregressive models, DecompDiff generates molecular coordinates and atom types jointly in a non-autoregressive and rotation-equivariant fashion. It explicitly models protein-ligand interactions in 3D space using an equivariant GNN, achieving high-quality binding poses and supporting unsupervised affinity prediction via learned structural features.

\textbf{Decomp-O}~\cite{katigbak2020alphaspace}: This method leverages AlphaSpace 2.0 for pocket representation and ligand design by introducing $\beta$-clusters—pseudomolecular constructs derived from concave protein surfaces. These $\beta$-clusters mimic the shape and atomic arrangement of ideal molecular binders, enabling direct pocket-ligand shape comparison. Decomp-O uses docking-based metrics such as the $\beta$-score to quantify pocket ligandability and supports ensemble pocket mapping for protein alignment. In SBDD pipelines, Decomp-O often serves as a pocket-centric prior to guide structure-aware molecule generation, particularly effective when integrated with ligand optimization techniques.

\textbf{MolCRAFT}~\cite{qu2024molcraft}: MolCRAFT addresses key limitations of previous autoregressive and diffusion-based SBDD models, particularly mode collapse and hybrid discrete-continuous denoising. Instead of operating in hybrid space, it performs molecular generation entirely within a continuous parameter space, allowing smooth, noise-reduced updates and improved sample efficiency. The model integrates SE(3)-equivariance for geometry consistency and achieves state-of-the-art Vina scores and conformation stability under controlled molecular sizes. Extensive experiments show MolCRAFT produces chemically feasible, high-affinity ligands with minimal sampling steps, establishing it as one of the most robust SBDD baselines to date.

\subsection{Our Method: Chem3DLLM}
We compare against all baselines using the proposed Chem3DLLM framework. Our model jointly supports MCG and SBDD tasks through unified reversible compression, cross-modal integration, and chemical-reward-driven reinforcement learning. Detailed architectural components and training strategies are described in Section ``\textbf{3 Method}" of the main paper.

\begin{figure}[h!]
\centering
\includegraphics[width=0.98\columnwidth]{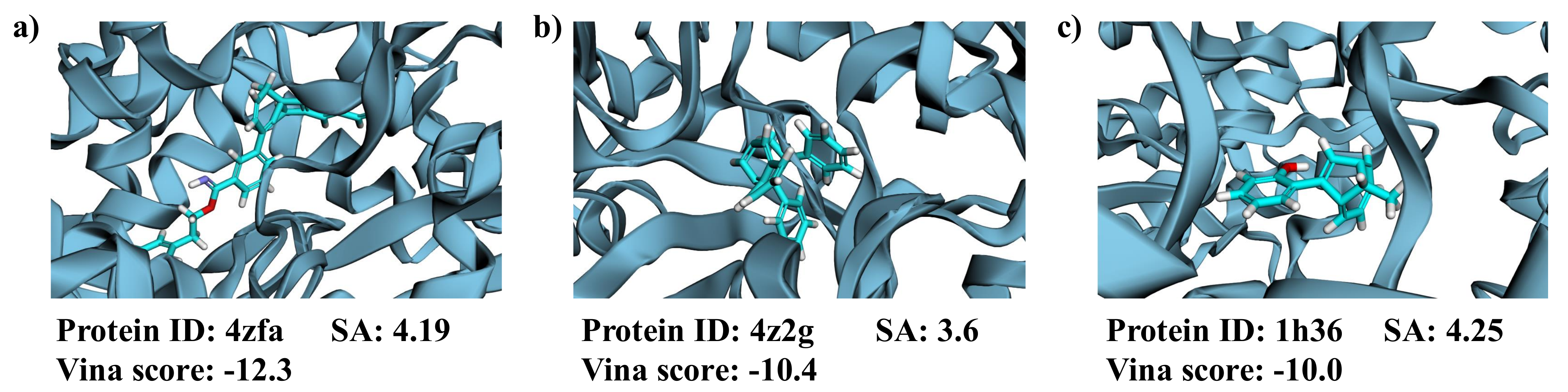} 
\caption{Additional case studies on structure-based drug design task. Chem3DLLM generates molecules that tightly fit the target protein pockets and yield favorable docking scores and SA values.}
\label{fig:case_SBDD}
\end{figure}

\begin{figure}[h!]
\centering
\includegraphics[width=\columnwidth]{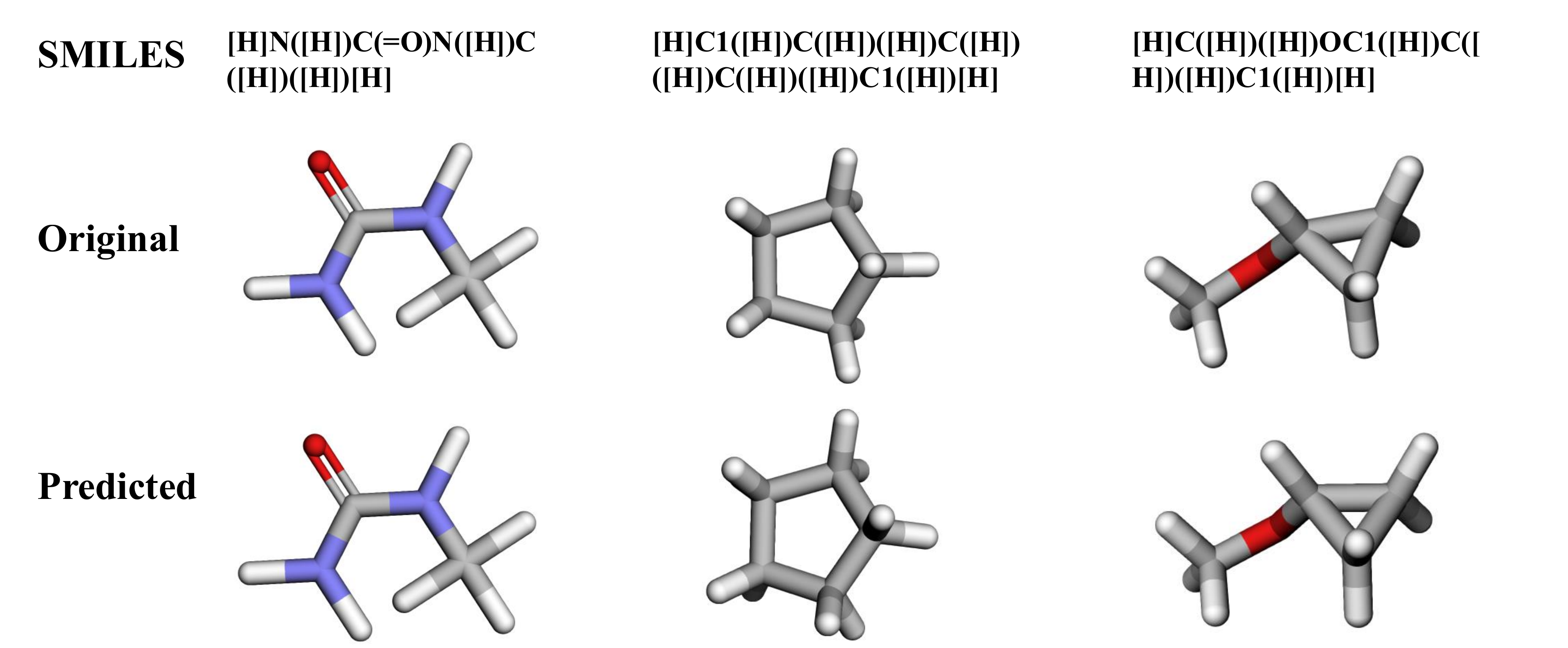} 
\caption{Additional case studies on molecular conformation generation task. Chem3DLLM accurately recovers the 3D atomic arrangements from SMILES across diverse structural motifs.}
\label{fig:case1_MCG}
\end{figure}

\section{Additional Visualization Examples}
\label{appendix:Additional_Visualization_Examples}

To further evaluate the effectiveness of Chem3DLLM in both conformation generation and structure-based drug design (SBDD) tasks, we present additional qualitative results beyond the main figures reported in Section ``\textbf{4 Experiments}.'' These examples are intended to illustrate the structural fidelity and chemical plausibility of generated molecules across more diverse inputs and conditions.

\subsection{Structure-Based Drug Design (SBDD)}

Figure~\ref{fig:case_SBDD} shows additional protein-ligand binding cases, where Chem3DLLM generates drug-like candidates conditioned on protein pocket geometry. The visualizations highlight the spatial compatibility of generated molecules with target binding sites. For example, the molecule generated for protein 4zfa achieves a Vina score of $-12.3$ and an SA (Synthetic Accessibility) score of 4.19, indicating strong predicted binding affinity with acceptable synthesis difficulty. Other cases such as 4z2g and 1h36 also demonstrate similarly tight docking poses, with all Vina scores below $-10.0$ and SA scores within a practical range ($< 5$). These results suggest that Chem3DLLM can produce pharmacologically relevant ligands with structural compatibility and synthetic realism.

\subsection{Molecular Conformation Generation (MCG)}

In Figure~\ref{fig:case1_MCG}, we present additional examples of conformation prediction from SMILES inputs. For each molecule, the top row shows the ground truth 3D structure (extracted from QM9), while the bottom row shows the structure predicted by Chem3DLLM. Despite the absence of explicit atom-level supervision during RL optimization, the predicted conformations preserve key chemical and geometric features—such as bond lengths, aromaticity, and stereochemistry—demonstrating high fidelity to ground-truth structures. These results further support the model’s ability to reconstruct fine-grained 3D arrangements using only compact tokenized inputs and molecular priors.

\section{Case Study: SDF File Compression}
\label{appendix:compression_case}

\begin{figure}[ht]
\centering
\begin{tcolorbox}[colback=gray!5!white, colframe=black!40, title=Raw SDF Content (RDKit 3D)]
\begin{Verbatim}[fontsize=\small]
RDKit          3D

 10  9  0  0  0  0  0  0  0  0999 V2000
   -2.9010   12.7890  -16.4760 C   0  0  0  0  0  0  0  0  0  0  0  0
   -3.6540   13.0410  -15.5400 O   0  0  0  0  0  0  0  0  0  0  0  0
   -2.7620   11.3220  -16.8420 C   0  0  0  0  0  0  0  0  0  0  0  0
   -1.7390   10.9930  -17.5170 O   0  0  0  0  0  0  0  0  0  0  0  0
   -3.6980   10.5520  -16.4260 O   0  0  0  0  0  0  0  0  0  0  0  0
   -1.8410   14.9530  -16.4260 C   0  0  0  0  0  0  0  0  0  0  0  0
   -2.2290   13.7370  -17.1450 N   0  0  0  0  0  0  0  0  0  0  0  0
   -1.3860   16.1600  -17.2870 C   0  0  0  0  0  0  0  0  0  0  0  0
   -2.2170   17.1100  -17.3780 O   0  0  0  0  0  0  0  0  0  0  0  0
   -0.2300   16.1240  -17.7680 O   0  0  0  0  0  0  0  0  0  0  0  0
  1  3  1  0
  1  7  1  0
  1  2  2  0
  3  4  2  0
  3  5  1  0
  6  8  1  0
  6  7  2  3
  8  9  2  0
  8 10  1  0
M  END
$$$$
\end{Verbatim}
\end{tcolorbox}
\caption{Raw 3D SDF format of a sample molecule. Each atom is represented by Cartesian coordinates and atomic type, and bond information is encoded in the bottom section.}
\label{fig:appendix_d_raw}
\end{figure}

\begin{figure}[h!]
\centering
\begin{tcolorbox}[
  colback=blue!3!white, 
  colframe=blue!50!black, 
  title=Compressed Format (Atoms + Bonds ), 
  sharp corners,
  boxrule=0.5pt,
  left=5pt, right=5pt, top=5pt, bottom=5pt
]
\begin{minipage}[c]{0.4\textwidth}
\centering
\ttfamily
C@-2.9010,12.7890,-16.4760 \\
O@-3.6540,13.0410,-15.5400 \\
C@-2.7620,11.3220,-16.8420 \\
O@-1.7390,10.9930,-17.5170 \\
O@-3.6980,10.5520,-16.4260 \\
C@-1.8410,14.9530,-16.4260 \\
N@-2.2290,13.7370,-17.1450 \\
C@-1.3860,16.1600,-17.2870 \\
O@-2.2170,17.1100,-17.3780 \\
O@-0.2300,16.1240,-17.7680 \\[0.8em]
\# 0-2:1 \quad 0-6:1 \quad 0-1:2 \\
\# 2-3:2 \quad 2-4:1 \quad 5-7:1 \\
\# 5-6:2 \quad 7-8:2 \quad 7-9:1
\end{minipage}%
\hfill
\begin{minipage}[c]{0.38\textwidth}
\centering
\includegraphics[width=\linewidth]{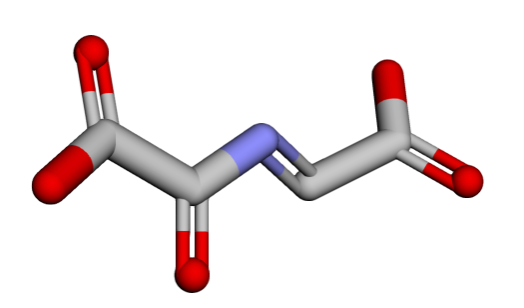}
\end{minipage}
\end{tcolorbox}
\caption{Compressed molecular representation produced by RCMT. Left: quantized atom coordinates and bond relations. Right: reconstructed 3D structure visualized for verification.}
\label{fig:appendix_d_comp}
\end{figure}

To illustrate the design and utility of our reversible compression scheme (RCMT), we present a case study comparing the raw SDF format and its compressed representation for a sample molecule. This example complements the algorithmic formulation in Section ``\textbf{Method}'' and visually demonstrates the mapping between the 3D structure and its compact tokenization.

\subsection{Original SDF Format}
Figure~\ref{fig:appendix_d_raw} displays the original molecular structure in standard RDKit SDF format. Each atom is specified by its Cartesian coordinates and atomic type, followed by a list of bonds with atom indices and bond orders. This format, while rich in detail, is verbose and not directly compatible with language models due to its numerical density and irregular structure.

\subsection{Compressed Representation and Visualization}
In Figure~\ref{fig:appendix_d_comp}, we show the output of our RCMT encoder applied to the same molecule. The top section lists quantized atomic entries in the form of "Element@x,y,z", while the bottom section encodes all non-zero bonds using a run-length compact format "i-j:o", where $i$ and $j$ are atom indices and $o$ is the bond order. This representation supports token-level alignment while retaining complete chemical and geometric fidelity. 

To validate structural consistency, we reconstruct the original 3D conformation from the compressed tokens and render the molecule in 3D. The geometry and connectivity are perfectly preserved (RMSD = 0), verifying the lossless nature of our reversible tokenizer. This case exemplifies how RCMT bridges the gap between chemically expressive 3D formats and discrete sequences suitable for language-based molecular modeling.

\end{document}